\documentclass[letterpaper,twocolumn,unpublished,11pt]{quantumarticle}
\pdfoutput=1
\usepackage[english]{babel}

\usepackage[T1]{fontenc}
\usepackage[utf8]{inputenc}
\IfFileExists{microtype.sty}{\usepackage{microtype}}{}

\usepackage[allcolors=quantumviolet,unicode=true,colorlinks,pdfauthor={Brandon Rodenburg and Keith
            Miller}]{hyperref}

\usepackage{graphicx}
\graphicspath{{figures/}}

\usepackage{lipsum}

\usepackage{amssymb,amsmath,braket,cancel}
\usepackage{units,siunitx}
\usepackage{commath} 

\DeclareMathOperator{\polylog}{polylog} 

\usepackage[square, numbers, sort&compress]{natbib}

\usepackage{ifpdf}

\usepackage{multirow}

\usepackage{framed}

\title{An Improved Volumetric Metric for Quantum Computers via more Representative Quantum Circuit Shapes.}
\author{Keith Miller}
\affiliation{Systems Engineering Innovation Center, MITRE, 7515 Colshire Drive, McLean, Virginia 22102, USA}
\affiliation{Homeland Security Systems Engineering and Development Institute (HSSEDI)™ operated by the MITRE Corporation}
\email{khmiller@mitre.org}
\author{Charles Broomfield}
\email{cbroomfield@mitre.org}
\affiliation{Homeland Security Systems Engineering and Development Institute (HSSEDI)™ operated by the MITRE Corporation}
\author{Ann Cox}
\affiliation{Department of Homeland Security, Science and Technology Directorate, USA}
\email{Ann.Cox@hq.dhs.gov}
\author{Joe Kinast}
\affiliation{Quantum Technologies Group, MITRE, 200 Forrestal Rd. Princeton, New Jersey 08540, USA}
\affiliation{Homeland Security Systems Engineering and Development Institute (HSSEDI)™ operated by the MITRE Corporation}
\email{jkinast@mitre.org}
\author{Brandon Rodenburg}
\affiliation{Quantum Technologies Group, MITRE, 200 Forrestal Rd. Princeton, New Jersey 08540, USA}
\affiliation{Homeland Security Systems Engineering and Development Institute (HSSEDI)™ operated by the MITRE Corporation}
\email{brodenburg@mitre.org}
\orcid{0000-0003-2150-0576}
\thanks{\smallskip\newline Approved for Public Release; Distribution Unlimited. Case
Number 22-0587 / DHS reference number 70RSAT22-021-01}

\begin{document}
\maketitle

\begin{abstract}
In this work, we propose a generalization of the current most widely used
quantum computing hardware metric known as the quantum volume~\cite{QV2018,
QV2019}. The quantum volume specifies a family of random test circuits defined
such that the logical circuit depth is equal to the total number of qubits used
in the computation. However, such square circuit shapes do not directly relate
to many specific applications for which one may wish to use a quantum computer.
Based on surveying available resource estimates for known quantum algorithms,
we generalize the quantum volume to a handful of representative circuit shapes,
which we call Quantum Volumetric Classes, based on the scaling behavior of the
logical circuit depth (time) with the problem size (qubit number). 

\end{abstract}
As a technology, quantum computing is in its infancy but developing rapidly. In
the near term, noisy and intermediate-scale quantum (NISQ) systems may become
useful for specific niche applications~\cite{PreskillNISQ2018}. In the long
term, with the development of fault-tolerant (FT) systems, this technology is
expected be extremely disruptive and transformative. Clear metrics to evaluate
this technology are crucial for evaluating and comparing performance of various
quantum devices and platforms in the near term, as well as creating
quantitative tools to better anticipate more long-term disruptions of this
technology.

An ideal metric has several key features, which we list in
\autoref{tab:IdealMetric} below. First, we want such a metric to be defined in
such a way that it is universal across potential quantum computers. This means
we want a metric that is defined at the logical computational level,
independent of the underlying physical platform. In addition, we want a metric
that is universal across maturity levels. We do not just want a metric for NISQ
devices, but one that applies equally to both NISQ and more mature FT systems
that utilize error correction. 

\begin{table}[htpb]
\begin{framed}
\begin{enumerate}
    \item Universal and platform independent
    \item Applicable to both near (NISQ) and long term (FT) systems
    \item Simple enough to be useful and understandable to non-experts
    \item Representative of the computational power needed to execute quantum
        algorithms
\end{enumerate}
\end{framed}
    \caption{Features of an ideal metric}
    \label{tab:IdealMetric}
\end{table}

We also require the metric to be simple and understandable enough to be useful,
including to non-experts. There will inevitably be trade-offs between precision
and simplicity. However, one of the key reasons for high-level metrics is to
provide some level of guidance to non-experts, whether they are end users,
application specialists, leadership within a company, or a government
organization. Such potential stakeholders need quantitative tools to make
strategic decisions or to track the technology over time.

Finally, the ideal metric needs to represent the computational power in a
straightforward way, i.e., a bigger number represents a more powerful device.
These values should also be tied closely to applications, making it possible to
more quickly determine if a device can run some specific application.
Conversely, having a metric closely aligned with the application space allows
specific applications to be framed directly in terms of this framework. This
makes any such system far more useful to potential end users.

The rest of this paper is laid out as follows.
\hyperref[sec:QV]{Section~\ref*{sec:QV}} gives a detailed description of the
quantum volume metric, including its shortcomings. In particular, there are two
fundamental issues. First, the quantum volume does not obviously relate to many
of the applications that one may wish to use a quantum computer for in terms of
resources needed to perform a computation. The second is the fact that the
quantum volume does not explicitly take into account quantum error correction,
a feature that will be necessary for this technology to mature beyond the NISQ
era that exists today. This is especially important in light of the rapid
progress towards implementing error correction below the fault-tolerant
threshold~\cite{QEC1, QEC2, QEC3, QEC4, QEC5, QEC6, QEC7, QEC8}.
\section{Quantum Volume}\label{sec:QV}
One of the most widespread high-level metrics currently in use is the quantum
volume, originally introduced by IBM in 2017~\cite{QV2018, QV2019}. The quantum
volume is meant to capture the ability of a quantum computing device to prepare
any state from a random sampling of the state space of a set of qubits. A
quantum computer with $n$ qubits is represented by a $2^n$-dimensional Hilbert
space. However, if the qubits are poorly controlled or subject to excess noise,
the device cannot effectively sample the full state space of the $n$ qubits.

\begin{figure}[htpb]
    \centering
    \includegraphics[width=\linewidth]{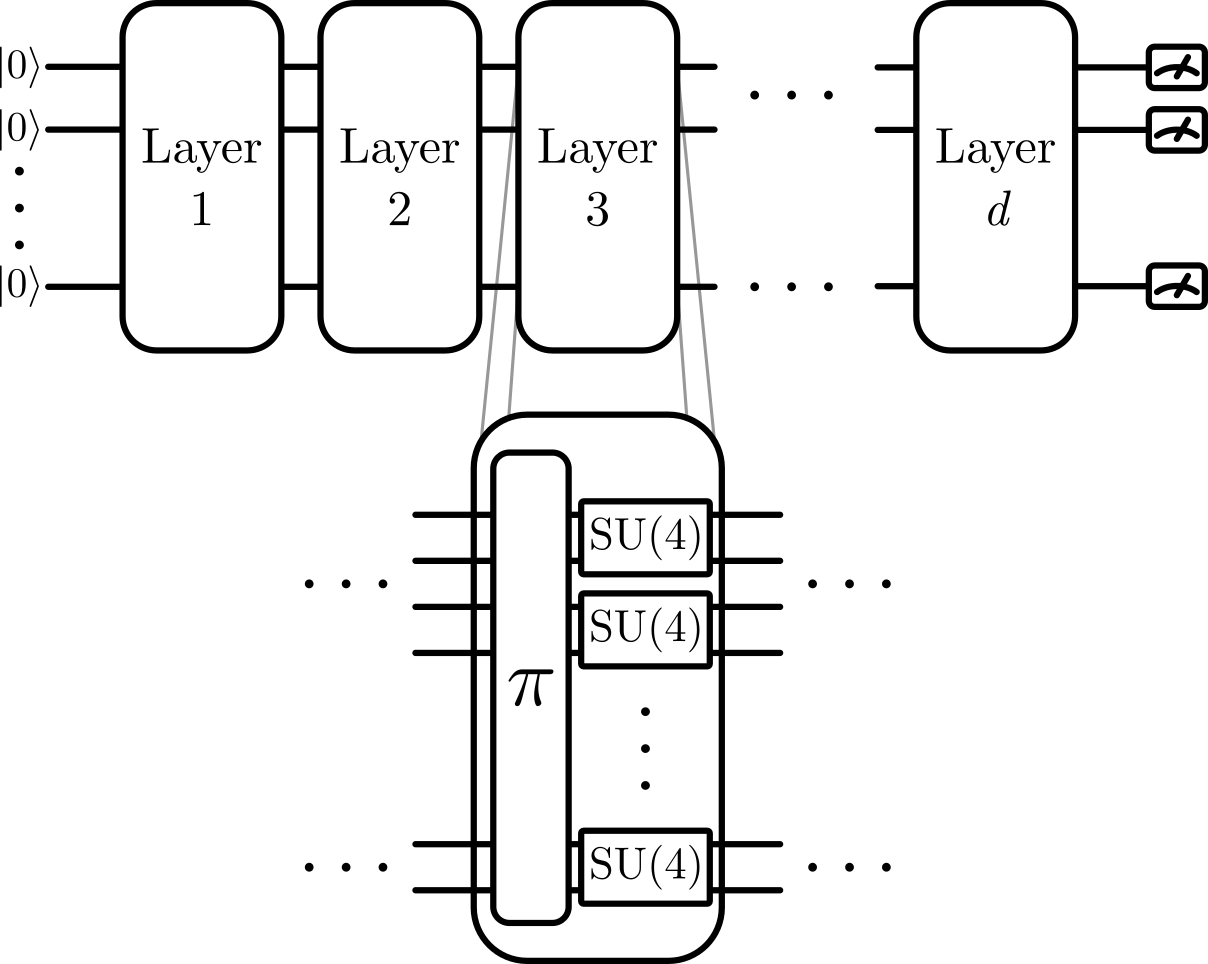}
    \caption{A circuit diagram for benchmarking the quantum volume. The circuit
    consists of $d$ layers of a random permutation of the qubits (represented
    by $\pi$) followed by random two-qubit SU(4) gates.}%
    \label{fig:QVCircuit}
\end{figure}

The ability to randomly access any part of the $2^n$-dimensional state space of
$n$ qubits, or equivalently the ability to randomly scramble an initial state,
requires a gate depth of $\mathcal O(n)$~\cite{FastScramble2008}. For this
reason, the quantum volume of a device was defined as the largest square
circuit that the device can implement. Mathematically this can be writted
as~\cite{QV2019}
\begin{equation}
    \log_2(V_Q) = \max_{n<n_\text{max}}\left(\min\left[n,d\right]\right),
    \label{eqn:QV}
\end{equation}
where $n$ is the subset of the $n_\text{max}$ qubits available used in the
circuit and $d$ is the circuit depth\footnote{The quantum volume was initially
defined as $V_Q = \max_n\left(\min\left[n,d\right]^2\right),$ representing the
space-time volume of a square $n\times n$ circuit~\cite{QV2018}. This was later
changed to the exponential definition in~\cite{QV2019}. However, due to the
exponential scaling of $V_Q$ with $n$, this number is usually quoted as
$\log_2(V_Q)$ instead.}. Each of the $d$ layers of the algorithm consists of a
random permutation of the qubits followed by the pairwise application of a
random 2-qubit SU(4) matrix (see \autoref{fig:QVCircuit}). The qubit
permutation may consist of sets of swap gates to move qubit states around, or a
simple logical relabeling of the qubits without applying any additional gates
to the extent the qubit layout and connectivity allows.

In order to determine the quantum volume for a specific hardware platform, a
definition of success is needed for a given test circuit. The original quantum
volume benchmark uses the heavy output criteria where so called ``heavy
outputs'' are measured with a probabily $>2/3$~\cite{QV2019,
HeavyOutputCriteria2017}. A downside of this criteria is that probability
outputs must be computed in advance, a problem that is generally
computationally hard classically. However, it is possible to construct
benchmark circuits that scale efficiently enough to be useful for large-scale
quantum circuits~\cite{MirrorCircuitBenchmarking2021}.

A problem with the quantum volume metric is the reliance on defining quantum
volume as an exponential, e.g. $2^n$, rather than just quoting the integer $n$
itself. This means that large-scale powerful systems will have exponentially
large values due to the fact that incremental improvements lead to large
differences in the quantum volume. Therefore, the quantum volume fails to be a
fair representation of computational power (item four in
\autoref{tab:IdealMetric}). For this reason, the logarithm of the quantum
volume is generally quoted rather than the volume itself. In this spirit, we
also adopt this convention when considering a generalized metric.

A second problem with the quantum volume is that as defined the quantum volume
does not explicitly specify how quantum error correction should be handled when
determining the metric value of a specific device, i.e., whether a quantum
volume test circuit is defined at the physical or logical level. As such, the
quantum volume is only applicable to the NISQ devices that currently exist
which do not utilize error correction. Therefore, this metric fails the
universality condition (item two) of \autoref{tab:IdealMetric}. Given our
emphasis on trying to relate computational power to known algorithms (e.g., by
looking for representative circuit shapes based on known algorithms), we choose
to define our metric at the logical level. By this we mean that circuit shapes
that define our metric are defined in terms of logical resources, and the
actual physical implementation of this circuit incorporates as much (or as
little) quantum correction overhead as needed to successfully implement the
circuit in question. Ideally, one would also like to have an idea of how the
computational resources defined at the logical level scale with the physical
features (e.g., qubit number and physical error rates), however, this type of
resource estimation is beyond the scope of this current work. 

A final problem that we see with the quantum volume is that a square $n\times
n$ circuit represents a minimum necessary circuit depth needed for many
potential quantum algorithms of interest (see \autoref{sec:QVShapes}). This
means that such a circuit represents at best the first step in a full
computation. In fact, many quantum algorithms have significantly greater
circuit depth than qubit number. That makes the quantum volume a poor
representative as a stand-in for the capability of the device (item four in
\autoref{tab:IdealMetric}).

A generalization of the metric that instead considers the qubit number and
circuit depth as separate independent resources has been
proposed~\cite{QVolumetricMetrics2020}. Although more robust, this
generalization comes at the cost of much greater complexity, as a quantum
computer is no longer represented by a single quantitative value, but rather an
entire family of successful circuit shapes and sizes. Therefore, this full
generalization although more descriptive and thus more useful as a generalized
framework for benchmarking~\cite{QEDC_AppBenchmarks2021}, fails our simplicity
criteria as an ideal metric by itself (item three in
\autoref{tab:IdealMetric}). 

In order to retain the advantages of this generalized
framework~\cite{QVolumetricMetrics2020} we look at a restricted subset of
possible circuit shapes. In this way we can balance the simplicity of a few
numbers representing our metric, while not restricting ourselves to the orginal
square circuit shape. To this end, a survey of resource requirements for known
quantum algorithms was performed in order to determine typical circuit shapes
to use in defining a generalized metric. The results of this survey and the
proposed circuit shapes are discussed in \autoref{sec:QVShapes}.
\section{Identifying Volumetric Shapes}\label{sec:QVShapes}
Quantum volume provides meaningful insight about a quantum computer’s ability
to implement algorithms which require a gate depth that scales no faster than
linearly in relation to number of qubits. In reality, however, several
algorithms appear to require a gate depth that scales faster than linearly in
relation to number of qubits. Thus, quantum volume has limited utility as a
universal metric for comparing the performance of quantum computers.

\begin{table*}[htpb]
\centering
\begin{tabular}{|l|ccc|}
\hline
\multirow{2}{*}{\textbf{Application Area}} & \multicolumn{3}{c|}{\textbf{Quantum Computing Era}}\\
\cline{2-4}
&\parbox[b]{0.7in}{\centering\textbf{NISQ}} & \parbox[b]{0.7in}{\centering\textbf{FT}} & \parbox[b]{0.7in}{\centering\textbf{All}}\\
\hline
Machine Learning            & 11   & 13 & 24                  \\
Optimization                & 9    & 5  & 14                  \\
Many-Body Physics/Chemistry & 9    & 7  & 16                  \\
Quantum Data Hiding         & 0    & 6  &  6                  \\
Numerical Solvers           & 0    & 3  &  3                  \\
Other                       & 0    & 2  &  2                  \\
\hline
\end{tabular}
\caption{Breakdown of the algorithms with known resource estimates considered
in this paper~\cite{
    QAlgRefID1, QAlgRefID2, QAlgRefID3, QAlgRefID4, QAlgRefID5, 
    QAlgRefID6, QAlgRefID7, QAlgRefID8, QAlgRefID9, QAlgRefID10, 
    QAlgRefID11, QAlgRefID12, QAlgRefID13, QAlgRefID14, QAlgRefID17,
    QAlgRefID34, QAlgRefID35, QAlgRefID36, QAlgRefID15, QAlgRefID16,
    QAlgRefID18, QAlgRefID19, QAlgRefID20, QAlgRefID21, QAlgRefID22,
    QAlgRefID23, QAlgRefID24, QAlgRefID25, QAlgRefID26, QAlgRefID27,
    QAlgRefID28, QAlgRefID29, QAlgRefID30, QAlgRefID31, QAlgRefID32,
    QAlgRefID33, QAlgRefID37, QAlgRefID38, QAlgRefID39, QAlgRefID40,
    QAlgRefID41, QAlgRefID42, QAlgRefID43, QAlgRefID44,
    QAlgRefID45}. We note that the total count of all algorithms exceeds 58
    due to the fact that some algorithms apply to both machine learning and
    optimization.}
\label{tab:AlgCounts}
\end{table*}

In order to look for more useful quantum circuit shapes, we performed a survey
of known quantum algorithms with available resource estimates. Data collection
included algorithms designed for both NISQ and FT devices. After reviewing over
225 research papers, we found 58~\cite{
    QAlgRefID1, QAlgRefID2, QAlgRefID3, QAlgRefID4, QAlgRefID5, 
    QAlgRefID6, QAlgRefID7, QAlgRefID8, QAlgRefID9, QAlgRefID10, 
    QAlgRefID11, QAlgRefID12, QAlgRefID13, QAlgRefID14, QAlgRefID17,
    QAlgRefID34, QAlgRefID35, QAlgRefID36, QAlgRefID15, QAlgRefID16,
    QAlgRefID18, QAlgRefID19, QAlgRefID20, QAlgRefID21, QAlgRefID22,
    QAlgRefID23, QAlgRefID24, QAlgRefID25, QAlgRefID26, QAlgRefID27,
    QAlgRefID28, QAlgRefID29, QAlgRefID30, QAlgRefID31, QAlgRefID32,
    QAlgRefID33, QAlgRefID37, QAlgRefID38, QAlgRefID39, QAlgRefID40,
    QAlgRefID41, QAlgRefID42, QAlgRefID43, QAlgRefID44, QAlgRefID45}
algorithms with resource estimates that could be used to approximate the
scaling of gate depth in relation to the number of qubits. Many of the
remaining research papers that were not used were due to a lack of clearly
defined circuit shape, e.g., by only specifying resources as exponential or
not.

Of the 58 algorithms considered, 27 algorithms were designed with NISQ
computers in mind~\cite{
    QAlgRefID1, QAlgRefID2, QAlgRefID3, QAlgRefID4, QAlgRefID5, 
    QAlgRefID6, QAlgRefID7, QAlgRefID8, QAlgRefID9, QAlgRefID10, 
    QAlgRefID11, QAlgRefID12, QAlgRefID13, QAlgRefID14, QAlgRefID17,
    QAlgRefID34, QAlgRefID35, QAlgRefID36},
    while 31 remain applicable only for larger or more universal fault-tolerant
    quantum computers~\cite{
    QAlgRefID15, QAlgRefID16, QAlgRefID18, QAlgRefID19, QAlgRefID20,
    QAlgRefID21, QAlgRefID22, QAlgRefID23, QAlgRefID24, QAlgRefID25,
    QAlgRefID26, QAlgRefID27, QAlgRefID28, QAlgRefID29, QAlgRefID30,
    QAlgRefID31, QAlgRefID32, QAlgRefID33, QAlgRefID37, QAlgRefID38,
    QAlgRefID39, QAlgRefID40, QAlgRefID41, QAlgRefID42, QAlgRefID43,
    QAlgRefID44, QAlgRefID45}. 
A breakdown of the number of algorithms by application area is given in
\autoref{tab:AlgCounts}.

\subsection{Assumptions During Data Collection}
Data collection required a number of assumptions. First, several research
papers provided resource estimates that could be used to approximate the
scaling of gate depth, but did not clearly or explicitly specify the scaling of
qubits. In these cases, qubits were assumed to scale linearly.

Second, the estimated scaling of gate depth for many of the algorithms depended
on other variables in addition to simple qubit count. For instance, on the
condition number of a specific matrix or function that is being computed, or
the desired error tolerance in a quantum simulation. In these situations, these
additional variables were treated as constants with the assumption that their
exact values would be determined when applied to a specific problem.

Finally, we found that some of the resource estimation papers provided specific
estimates for the scaling of gate depth, while others provided estimates for
the scaling of similar (but different) variables such as gate count,
operations, runtime, or time complexity (See \autoref{tab:DepthEstimationType}
for a breakdown of the algroithms by depth estimation type). Estimates for the
scaling of gate depth, runtime, and time complexity typically account for the
ability to implement quantum gates in parallel whereas estimates for the
scaling of gate count and operations typically do not. Therefore, estimates
associated with gate count or operations were considered overestimations that
needed to be accounted for when analyzing the data. 

\begin{table}[htpb]
\begin{tabular}{|c|c|}
    \hline
    \textbf{Circuit Depth Estimate Type} & \textbf{Count}\\
    \hline
    Gate Depth & 14\\
    Gate Count/Operations & 19\\
    Runtime/Time Complexity/etc. & 25\\
    \hline
\end{tabular}
\centering
\caption{Count of how many algorithms fall within each category of gate depth estimation types.}
\label{tab:DepthEstimationType}
\end{table}

\subsection{Categorization of Circuit Shapes}
\begin{figure*}[htpb]
    \centering
    \includegraphics[width=0.75\linewidth]{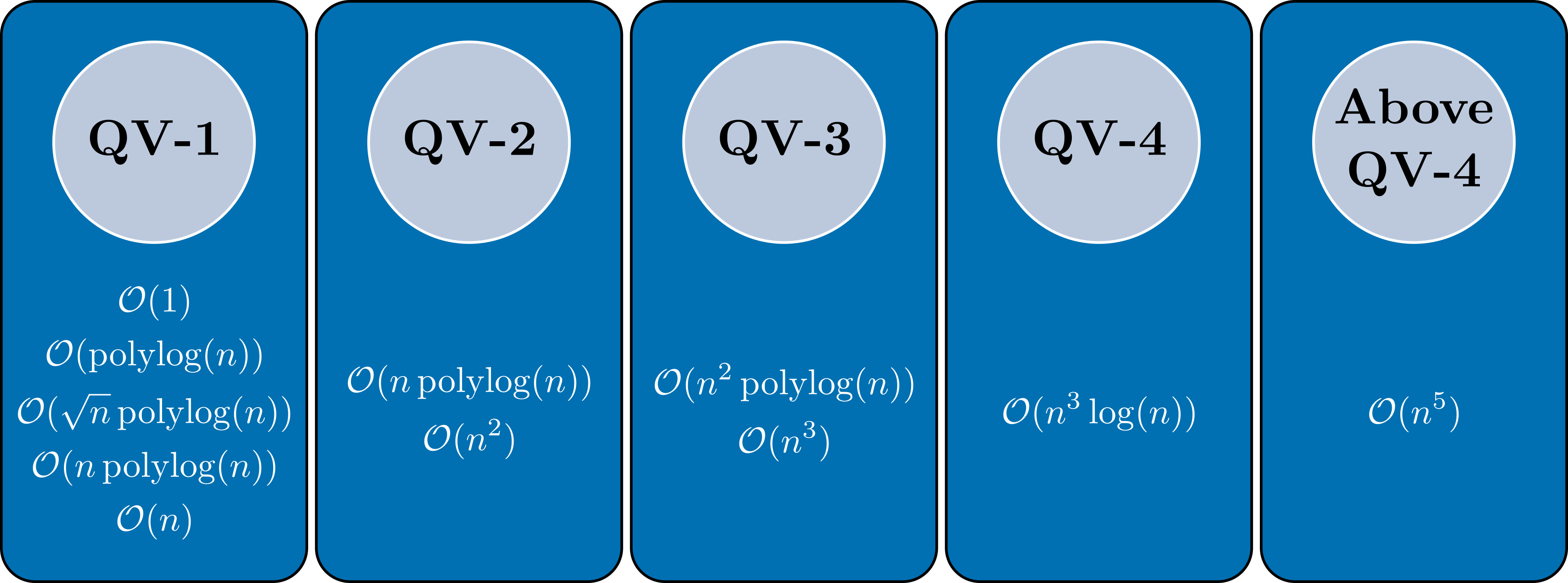}
    \caption{The Quantum Volumetric Classes used in this work, as well as the
        initial assumptions about which circuit depths from
        \autoref{eqn:PossibleDepths} fall within each class.}
    \label{fig:QVClassAlignment}
\end{figure*}

Data analysis first required identifying the circuit shapes that align with
each algorithm’s specific scaling of gate depth. We considered each algorithm's
circuit shape, $\mathcal S$, to be of the form
\begin{equation}
    \mathcal S = n \times d(n),
\end{equation}
where the circuit depth $d(n)$ is explicitly some function of the number of
required qubits $n$. The scaling of circuit depth for all algorithms fell into
one of the following forms, arranged from slower-growing to faster-growing:
\begin{equation}
d(n)\sim
\begin{cases}
   \mathcal O(1)\\
    \mathcal O(\polylog(n))\\
    \mathcal O(\sqrt{n}\polylog(n))\\
    \mathcal O(n\polylog(n))\\
    \mathcal O(n)\\
    \mathcal O(n\polylog(n))\\
    \mathcal O(n^2)\\
    \mathcal O(n^2\polylog(n))\\
    \mathcal O(n^3)\\
    \mathcal O(n^3\log(n))\\
    \text{or}\\
    \mathcal O(n^5)
\end{cases}.
\label{eqn:PossibleDepths}
\end{equation}

Ignoring $\polylog(n)$ factors in \autoref{eqn:PossibleDepths}, we filtered the
possible circuit shapes into the following ``Quantum Volumetric Classes''
listed in \autoref{fig:QVClassAlignment}. Most resource estimates for scaling
are rough order of magnitude estimates and ignore things such as overheads
which scale at least of order $\log(n)$~\cite{QCDepthOverhead2020}. Therefore,
we end up with a more reasonable number of categories with minimal loss of
informational value.

Mathematically, these classes can be written as
\begin{equation}
\begin{split}
    \text{QV-1} &= \max_{n<n_\text{max}}\Big(\min\big[n,d      \big]\Big)\\
    \text{QV-2} &= \max_{n<n_\text{max}}\Big(\min\big[n,d^{1/2}\big]\Big)\\
    \text{QV-3} &= \max_{n<n_\text{max}}\Big(\min\big[n,d^{1/3}\big]\Big)\\
    \text{QV-4} &= \max_{n<n_\text{max}}\Big(\min\big[n,d^{1/4}\big]\Big),
\end{split}
\label{eqn:QVClasses}
\end{equation}
where the variables are defined just as they were in \autoref{eqn:QV}. It
should be noted that we do not follow the exponential convention from the
original quantum volume for our classes. In particular, although QV-1 contains
the exact same information as the original quantum volume $V_Q$, these two are
not equal. Instead the relationship between the two is given as
\begin{equation}
    \text{QV-1} \equiv \log_2(V_Q).
\end{equation}
This should not present an issue as it is already common to quote $\log_2(V_Q)$
rather than $V_Q$ anyways.

As an example of how these classes work, a given quantum computer may score a
value of $n$ for the QV-1 metric. This corresponds to the determination that
the device is able to implement square circuits up to a size $\mathcal
S=n\times n,$ and is equivalent in this case to a quantum volume of
$\log_2(Q_V)$. Likewise a device with a QV-2 score of $n$ means the device can
successfully implement circuits whose depth grows quadratically with qubit
number up to a circuit shape of $\mathcal S=n\times n^2$. In general, a value
of $n$ for the QV-$k^\text{th}$ volumetric metric corresponds to the circuit of
shape $\mathcal S = n\times n^k.$ 

The following subsections walk through the data analysis process using these
classes of quantum volume metrics. First we will present initial results
without any adjustments to the data. Then, we will highlight adjustments made
to the data based on (1) an assumption that algorithms will require additional
gate depth to account for gate overhead and (2) the lack of parallelism in
resource estimates for the scaling of gate count and operations. The final
subsection then presents final results.

\section{Results}
The following subsections walk through the data analysis process using these
classes of quantum volume metrics. In \autoref{sec:InitialResults}, we will
present initial results without any adjustments to the data. Then,
\autoref{sec:ResultAdjustments} will highlight adjustments made to the data
based on (1) an assumption that algorithms will require additional gate depth
to account for gate overhead and (2) the lack of parallelism in resource
estimates for the scaling of gate count and operations.
\hyperref[sec:FinalResults]{Section~\ref*{sec:FinalResults}} then presents
final results.

\subsection{Initial Results}\label{sec:InitialResults}
\begin{table}[htpb]
\centering
\begin{tabular}{|c|c|c|}
\hline
\textbf{Class} & \textbf{Circuit Depth} & \textbf{Counts}\\
\hline
\multirow{5}{*}{QV-1}
& $\mathcal O(n)$                   & 16 (28\%) \\
& $\mathcal O(\log(n))$             &  8 (14\%) \\
& $\mathcal O(\log^x(n))$           &  6 (10\%) \\
& $\mathcal O(1)$                   &  2 (3\%)  \\
& $\mathcal O(\sqrt{n}\polylog(n))$ &  1 (2\%)  \\
\hline
\multirow{2}{*}{QV-2}
& $\mathcal O(n^2)$                 & 10 (17\%) \\
& $\mathcal O(n\polylog(n))$        &  2 (3\%)  \\
\hline
\multirow{2}{*}{QV-3}
& $\mathcal O(n^3)$                 &  8 (14\%) \\
& $\mathcal O(n^2\polylog(n))$      &  1 (2\%)  \\
\hline
QV-4
& $\mathcal O(n^3\log(n))$          &  2 (3\%)  \\
\hline
QV-5
& $\mathcal O(n^5)$                 &  2 (3\%)  \\
\hline
\end{tabular}
\caption{Number of algorithms by circuit depth scaling and grouped by QV class.}
\label{tab:InitialResultCounts}
\end{table}

Our initial sorting of algorithms into QV classes is shown in
\autoref{fig:QVClassAlignment}. The counts for the number of algorithms for
each of the given depth scalings of \autoref{eqn:PossibleDepths} is given in
\autoref{tab:InitialResultCounts}, together with the initial sorting by QV
class. Under this initial grouping, 33 (57\%) algorithms fall into QV-1, 12
(21\%) into QV-2, 9 (16\%) into QV-3, and 2 (3\%) into either QV-4 or QV-5.
Therefore the original quantum volume, represented by our QV-1 class, accounts
for only $57\%$ of the quantum algorithms surveyed. If we add the additional
categories of QV-2, QV-3, and QV-4 to the original quantum volume metric, we
would increase coverage to $93\%$ of algorithms surveyed.

\begin{figure}[htpb]
    \centering
    \includegraphics[width=\linewidth]{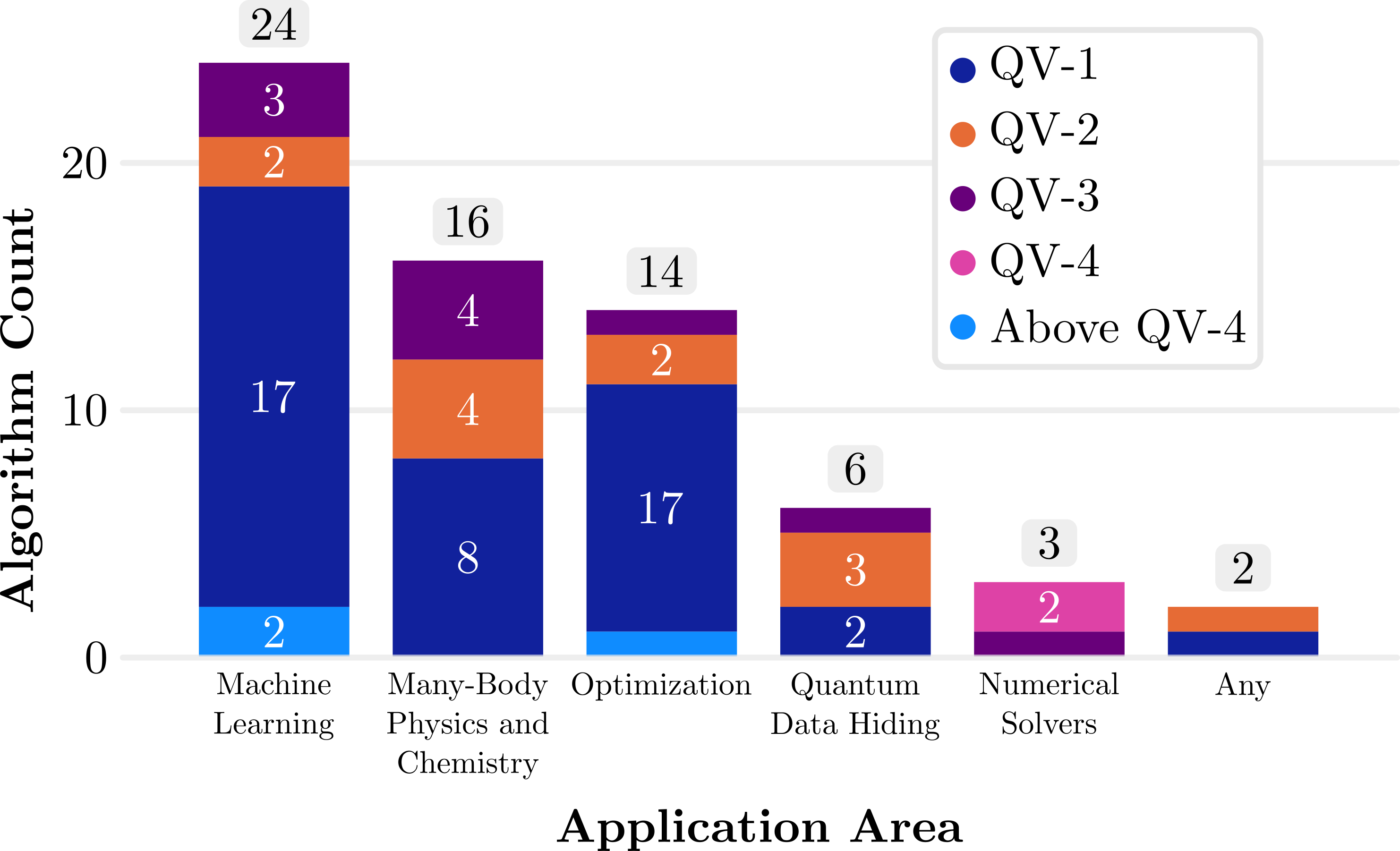}
    \caption{Initial breakdown of application areas by QV class.}
    \label{fig:InitialResults}
\end{figure}

When analyzing the initial data by application area we notice a few trends (see
\autoref{fig:InitialResults}). First, machine learning and optimization
algorithms primarily seem to align with QV-1. Many-body physics and chemistry
algorithms align evenly across QV-1 and the combination of QV-2 and QV-3.
Quantum data hiding algorithms align with QV-1 for watermarking and QV-2/QV-3
for steganography. Numerical solvers (i.e., Shor’s algorithm) align with QV-3
or QV-4.

\subsection{Adjustments to Results}\label{sec:ResultAdjustments}
After an initial analysis, the categorization of algorithms into their
respective QV classes was adjusted to account for the following assumptions.
First, it is anticipated that implementing algorithms on quantum computers will
require gate overhead due to the physical limitations of qubit connectivity
within the physical devices. Qubits will likely not be able to directly
interact with all other qubits; therefore, additional quantum gates will likely
be needed to make the necessary connections. This gate overhead is expected to
be at least of order $\log(n)$~\cite{QCDepthOverhead2020}. 

\begin{figure*}[htpb]
\centering
\includegraphics[width=0.75\linewidth]{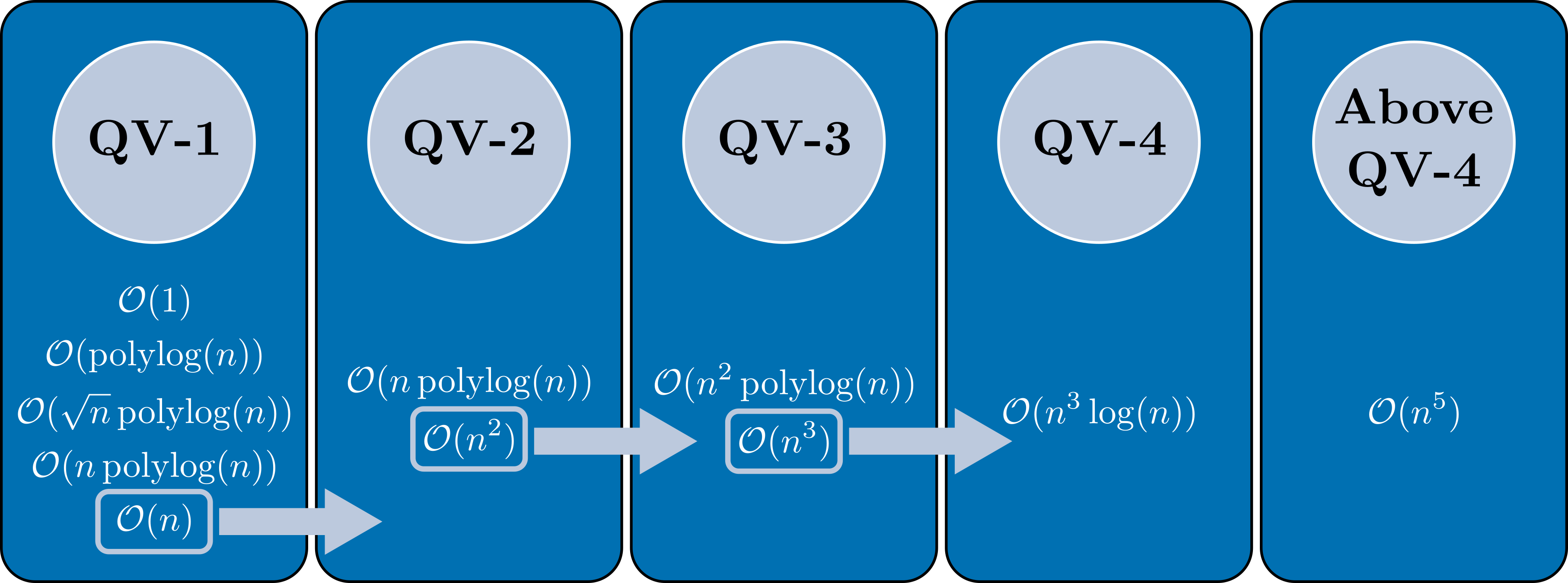}
\caption{Adjustments to the categorization of algorithms into QV classes as
    origionally represented in \autoref{fig:QVClassAlignment} for different
    scaling behaviors for the circuit depths.} 
\label{fig:QVClassAlignmentAdjusted}
\end{figure*}

Keeping this overhead in mind, the initial data presented in
\autoref{sec:InitialResults} is likely an underestimation of actual gate depth
requirements. To account for this, algorithms with circuit depth estimates
scaling exactly with the shape of their respective QV metrics are pushing the
upper limits of the metric, and they should therefore move to the next higher
QV metric to leave room for potential gate overhead. To adjust for this,
algorithms with circuit depths $\mathcal O(n)$ were moved from QV-1 to QV-2,
$\mathcal O(n^2)$ were moved into QV-3, and $\mathcal O(n^3)$ were considered
part of QV-4. This change is represented graphically in
\autoref{fig:QVClassAlignmentAdjusted}. 

\begin{table}[htpb]
\centering
\begin{tabular}{|c|c|c|}
\hline
\textbf{Depth} & \textbf{Adjust Class} & \textbf{Keep Initial Class}\\
\hline
$\mathcal O(n)$   & 12 & 4\\
$\mathcal O(n^2)$ &  6 & 4\\
$\mathcal O(n^3)$ &  4 & 4\\
\hline
\end{tabular}
\caption{A count of how many algorithms for borderline cases were adjusted to
    the next higher QV class according to method shown in
    \autoref{fig:QVClassAlignmentAdjusted}. This assessment was based on which
    circuit depth estimation type was used (see
    \autoref{tab:DepthEstimationType}). Estimations based on total gate counts
    were kept at their initial QV class, while all others were adjusted.}
\label{tab:AdjustmentCount}
\end{table}

A second consideration was the fact that the resource estimates of the various
algorithms were not all of the same type (see
\autoref{tab:DepthEstimationType}). In particular, resource estimations that
rely on assumptions of total gate or operation counts do not take into account
parallelisation of the actual algorithm. Since these cases already represent
overestimations of gate depth, we did not move any such algorithm to a higher
QV class. Counts for how many algorithms were moved to a higher QV classes is
shown in \autoref{tab:AdjustmentCount}.

\subsection{Final Results}\label{sec:FinalResults}
\begin{table}[htpb]
\centering
\begin{tabular}{|c|c|c|}
\hline
\textbf{Class} & \textbf{Circuit Depth} & \textbf{Counts}\\
\hline
\multirow{5}{*}{QV-1}
& $\mathcal O(n)$                   &  4 (7\%) \\
& $\mathcal O(\log(n))$             &  8 (14\%) \\
& $\mathcal O(\log^x(n))$           &  6 (10\%) \\
& $\mathcal O(1)$                   &  2 (3\%)  \\
& $\mathcal O(\sqrt{n}\polylog(n))$ &  1 (2\%)  \\
\hline
\multirow{3}{*}{QV-2}
& $\mathcal O(n)$                   & 12 (21\%) \\
& $\mathcal O(n^2)$                 &  4 (7\%) \\
& $\mathcal O(n\polylog(n))$        &  2 (3\%)  \\
\hline
\multirow{3}{*}{QV-3}
& $\mathcal O(n^2)$                 &  6 (10\%) \\
& $\mathcal O(n^3)$                 &  4 (7\%) \\
& $\mathcal O(n^2\polylog(n))$      &  1 (2\%)  \\
\hline
\multirow{2}{*}{QV-4}
& $\mathcal O(n^3)$                 &  4 (7\%) \\
& $\mathcal O(n^3\log(n))$          &  2 (3\%)  \\
\hline
QV-5
& $\mathcal O(n^5)$                 &  2 (3\%)  \\
\hline
\end{tabular}
\caption{Number of algorithms by circuit depth scaling and grouped by QV class
based on the adjustments described in \autoref{sec:ResultAdjustments}.}
\label{tab:FinalResults}
\end{table}

After adjusting the data as described in \autoref{sec:ResultAdjustments} we
arrived at the final results, presented in \autoref{tab:FinalResults} below.
Under this adjusted grouping, the utility of QV-1 is reduced, as it now only
accounts for 21 (36\%) of the algorithms, while QV-2 saw a significant boost to
18 (31\%) algorithms. QV-3 increased slightly to 11 (19\%), while QV-4 now has
6 (10\%) of all algorithms surveyed. QV-5 remained at 2 (3\%). Therefore we
see, a set of metrics consisting of QV-1, QV-2, QV-3, and QV-4 should cover a
vast majority of quantum algorithms ($91\%$).

\begin{figure}[htpb]
    \centering
    \includegraphics[width=\linewidth]{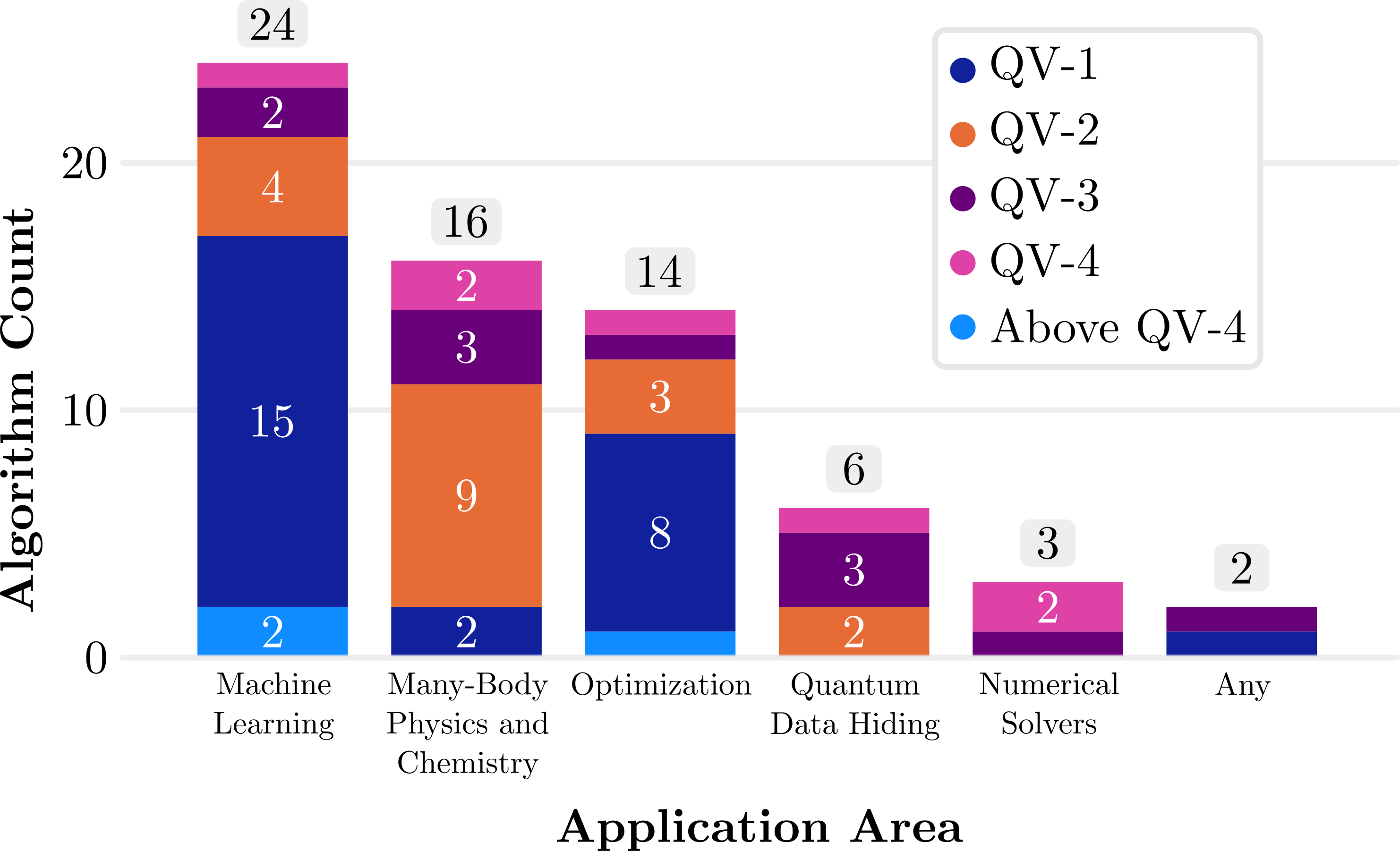}
    \caption{Adjusted breakdown of application areas by QV class.}
    \label{fig:FinalResults}
\end{figure}

We again analyzed the updated data by application area, which we represent
graphically in \autoref{fig:FinalResults} Machine learning and optimization
algorithms continue to align most strongly with QV-1. Many-body physics and
chemistry algorithms now align most strongly with QV-2. Although machine
learning, optimization, and many-body physics and chemistry algorithms align
most strongly to a specific QV metric, they each have at least one
implementation for each QV metric. Quantum data hiding algorithms now align
with QV-2 for watermarking and QV-3/QV-4 for steganography. Numerical solvers
(i.e. Shor’s algorithm) continue to align with QV-3 or QV-4.

\section{Conclusions}
Quantum computing is a rapidly developing technology that promises to be
broadly disruptive across many fields. This necessitates the development of
better metrics to enable potential end users to evaluate and compare hardware
platforms for their desired applications, as well as to track the development
of the technology over time. The generalized quantum volume classes presented
in this work (see \autoref{eqn:QVClasses}) provide a generalization to the
original quantum volume benchmark that is applicable to a much broader range of
potential applications, while not becoming too complex to be useful to the
potential end users for these applications.

Adoption of these metrics by the community would help in a number of ways.
These metrics more explicitly tie the power of quantum computing to specific
use cases, enhancing the utility to end-users. Focusing on developing metrics
from the end-user perspective will also help companies, governments, and other
entities to make more informed decisions concerning this field. Adoption of
these metrics or similar frameworks will help guide the field by quantifying
how success and improvement is measured. Having a consistent framework should
also help minimize some of the assumptions specified in \autoref{sec:QVShapes}.

In the coming years we can expect to see more powerful quantum computing
systems with both more qubits and better fidelities. Soon, we will even see
systems whose error rates fall below error correction thresholds, opening the
door to further progress through the application of quantum error correction to
these systems. The original quantum volume does not specify how to account for
error correction, i.e., whether the specified square circuit shape applies at
the logical or physical level. In order for a metric to be sufficiently simple
and useful to end users who may not be experts in quantum computing, the metric
should be defined at the logical level to best align with the application
space. However, it would also be useful to be able to connect the user space
metric, such as that presented here, with estimates of system level
performance. Such an accounting is beyond the scope of this work, but is
forthcoming in a future paper.

\section*{Acknowledgements}

The Homeland Security Act of 2002 (Section 305 of PL 107-296, as codified in 6
U.S.C.~185), herein referred to as the ``Act,'' authorizes the Secretary of the
Department of Homeland Security (DHS), acting through the Under Secretary for
Science and Technology, to establish one or more federally funded research and
development centers (FFRDCs) to provide independent analysis of homeland
security issues. MITRE Corp.~operates the Homeland Security Systems Engineering
and Development Institute (HSSEDI) as an FFRDC for DHS under contract
70RSAT20D00000001.

The HSSEDI FFRDC provides the government with the necessary systems engineering
and development expertise to conduct complex acquisition planning and
development; concept exploration, experimentation and evaluation; information
technology, communications and cyber security processes, standards,
methodologies and protocols; systems architecture and integration; quality and
performance review, best practices and performance measures and metrics; and,
independent test and evaluation activities. The HSSEDI FFRDC also works with
and supports other federal, state, local, tribal, public and private sector
organizations that make up the homeland security enterprise. The HSSEDI FFRDC’s
research is undertaken by mutual consent with DHS and is organized as a set of
discrete tasks. This report presents the results of research and analysis
conducted under: 70RSAT22FR0000021, ``DHS Science and Technology Directorate
TCD QIS Capabilities.''

The results presented in this report do not necessarily reflect official DHS
opinion or policy.

\bibliographystyle{unsrtnat}
\bibliography{References.bib}

\end{document}